\documentclass[
aps,prd,
12pt,
nopreprintnumbers,
showpacs,
eqsecnum,
nofootinbib
]{revtex4-1}

\usepackage{graphicx}
\usepackage{amssymb}

\def\sm{\mbox{\footnotesize$\sum_l$}}
\def\pl{\mbox{$\frac{r_++r_-}{2}$}}
\def\mi{\mbox{$\frac{r_+-r_-}{2}$}}

\begin{document}

\title{Electric and dilatonic f{}ields of a charged massive particle at rest
in the f{}ield of a charged dilaton black hole}
\author{Nahomi Kan}\email[]{kan@gifu-nct.ac.jp}
\affiliation{National Institute of Technology, Gifu College,
Motosu-shi, Gifu 501-0495, Japan}
\author{Kiyoshi Shiraishi\footnote{Author to whom any
correspondence should be addressed.}}\email[]{shiraish@yamaguchi-u.ac.jp}
\affiliation{
Graduate School of Sciences and Technology for Innovation, Yamaguchi
University, Yamaguchi-shi, Yamaguchi 753--8512, Japan}
\date{\today}

\begin{abstract}
We study linear-perturbation equations for the two-body system of a charged
dilaton black hole, of which dilaton coupling constant is $\alpha$, and
a static particle with mass 
$m$, electric charge
$q$, and dilatonic charge
$\beta m$. We find that a consistent condition for the coupled equations
corresponds to the equilibrium condition of the test particle. The expressions of
classical fields are given in closed analytical
formulas in the most interesting case with $\beta=\alpha$. We examine the
electrical field around a charged dilaton black hole especially in
the limit of the maximum electric charge and we find the electric Meissner effect
which has been found for the Reissner--Nordstr\"om black hole in the
Einstein--Maxwell system.
\end{abstract}



\maketitle

\section{Introduction}
\label{introduction}
The electric field created by a point source around a fixed black hole background
has been widely discussed by many authors (see for example
\cite{HR1,LL,Hanni,LT,WL,LHR} and references therein). In addition, perturbative
analyses incorporating back reactions have already been carried out for the
system consisting of a static particle in the vicinity of a Reissner--Nordstr\"om
black hole of which analytical solutions are well known
\cite{BGR1,BGR2,BGR3,BGR4}. In this case, the electrical repulsive force attains
the equilibrium configuration of the black hole and the point charge.

In the low-energy limit of string theory and supergravity theories, which are
candidates for a unified theory including gravity, a massless scalar field
called a dilaton field appears. It is also known that the dilaton field is
non-minimally coupled to the Maxwell field with an exponential coupling. On the
other hand, dilaton fields also appear in the context of Kaluza--Klein theories.
The dilaton mediates scalar long-range forces, like gravity and electric forces.
Thus, the Einstein--Maxwell-dilaton theory is an interesting and simple
theoretical arena for exploring modified gravities which involve scalar and
vector fields, while black holes are very important objects in terms of strong
gravity.

In this paper, we approach the case of a static particle of mass $m$, charge
$q$ and dilaton charge $\beta m$, in the field of a static, spherically symmetric
 charged dilaton black hole \cite{GM,GHS,HW,Rakhmanov} in the
Einstein--Maxwell-dilaton system with the dilaton coupling constant $\alpha$. We
derive linear-perturbation equations for the metric field, electric field and
dilaton field from field equations of the system, and obtain them up to
first-order contributions such as mass and charges of a particle in
the present paper. In this case,
the attractive forces by the dilatonic and gravitational forces cancel the
repulsive forces by the electric field in the two-body system. 

The particle is assumed to be at rest at the point $\textbf{x}=\textbf{b}$.
The non-vanishing component of the energy-momentum tensor, the electric current
density, and the dilaton charge of the particle are given by%
\footnote{Exact expressions are provided later in Sec.~\ref{sec3}.}
\begin{equation}
T^p_{00}\propto m \delta^3(\mathbf{x}-\mathbf{b})\,,\quad
J^0_p\propto q \delta^3(\mathbf{x}-\mathbf{b})\,,\quad
\Sigma^p\propto \beta m \delta^3(\mathbf{x}-\mathbf{b})\,,
\end{equation}
and the other components are zero.
These will be put into the Einstein equations, the electromagnetic field
equations, and the dilaton equations all combined.
As an important preceding study, the electric field surrounding a
Reissner--Nordstr\"om black hole was investigated by Bini, Geralico and Ruffini
\cite{BGR1,BGR2,BGR3,BGR4} for a point charge at rest. We follow their method and
proceed our analysis on the charged dilaton black hole.

The structure of the present paper is as follows.
In the next section, 
the exact solution for a spherical charged dilaton black hole in the
Einstein--Maxwell-dilaton gravity is reviewed.
We consider linear perturbative fields about the charged dilaton black hole
solution and derive the differential equations for them in Sec.~\ref{sec3}.
We also clarify the consistency condition for the particle at rest in this
section.  
Section \ref{sec4} is devoted to the simple but interesting case with
$\beta=\alpha$, and we give a closed form of perturbed fields in this case. In
Sec.~\ref{sec5}, the electric and dilaton fields are explored
 for the $\beta=\alpha$ case. The electric Meissner effect
\cite{BGR1,BGR2,BGR3,BGR4} can be found in the limit of the maximally-charged
black hole. We compare the case of the maximally-charged dilaton black hole with
the well-known exact multi-black hole solution in the Einstein--Maxwell-dilaton
system in Sec.~\ref{sec6}, and we confirm the coincidence of them up to the
linear order of the small mass of the particle. We analytically study the system
of the maximally-charged dilaton black hole and the particle for
$\beta\ne\alpha$ in Sec.~\ref{sec7}. We summarize our results in
Sec.~\ref{conclusion}.

We use the metric convention $(-+++)$ and adopt $G=c=1$ units (where $G$ is the
Newton constant and $c$ is the light speed) throughout the present paper.

\section{The electrically charged dilaton black hole}
\label{sec2}

In this section, we give a brief introduction to the charged dilaton black hole
in $1+3$ dimensions. We consider theories including coupled gravitational,
electromagnetic, and dilaton fields with the action:
\begin{equation}
S=\int d^4x\frac{\sqrt{-g}}{16\pi}\left[
R-2(\nabla\Phi)^2-e^{-2\alpha\Phi}F^2
\label{ac}
\right]\,,
\end{equation}
where $R$ is the scalar curvature, $\Phi$ is a real scalar field (the dilaton)
and $(\nabla\Phi)^2\equiv g^{\mu\nu}\nabla_\mu\Phi\nabla_\nu\Phi$,
while the electromagnetic field strength is defined by $F_{\mu\nu}=\partial_\mu
A_\nu-\partial_\nu A_\mu$ with the $U(1)$ gauge field $A_\mu$, and $F^2\equiv
F_{\mu\nu}F^{\mu\nu}$. The constant $\alpha$ represents the dilaton coupling.
When $\alpha=0$, the action reduces to a usual Einstein--Maxwell action.

The field equations derived from the action (\ref{ac}) without external
sources are
\begin{eqnarray}
& &\nabla_\nu(e^{-2\alpha\Phi}F^{\mu\nu})=0\,,\\
& &\nabla^2\Phi+\frac{\alpha}{2}e^{-2\alpha\Phi}F^2=0\,,\\
& &G_{\mu\nu}\equiv R_{\mu\nu}-\frac{1}{2}Rg_{\mu\nu}=8\pi T^\Phi_{\mu\nu}
+8\pi T^F_{\mu\nu}\,,
\end{eqnarray}
where
\begin{eqnarray}
8\pi T^F_{\mu\nu}&=&e^{-2\alpha\Phi}\left[2F_{\mu\lambda}F_\nu{}^\lambda
-\frac{1}{2}F^2g_{\mu\nu}\right]\,,
\label{TF}\\
8\pi T^\Phi_{\mu\nu}&=&2\nabla_\mu\Phi\nabla_\nu\Phi-(\nabla\Phi)^2g_{\mu\nu}\,.
\label{TPhi}
\end{eqnarray}

A static spherically symmetric solution of the equations represents an
electrically charged dilaton black hole with the line element
\cite{GM,GHS,HW,Rakhmanov}
\begin{equation}
ds^2=-f(r)dt^2+\frac{1}{f(r)}dr^2+r^2\sigma^2(r)(d\theta^2+\sin^2\theta
d\varphi^2)\,,
\end{equation}
where
\begin{equation}
f(r)=\left(1-\frac{r_+}{r}\right)
\left(1-\frac{r_-}{r}\right)^\frac{1-\alpha^2}{1+\alpha^2}\quad
\mbox{and}\quad
\sigma(r)=
\left(1-\frac{r_-}{r}\right)^\frac{\alpha^2}{1+\alpha^2}\,.
\end{equation}
For later convenience, we define a function of $r$,
\begin{equation}
\Delta(r)\equiv\left(1-\frac{r_+}{r}\right)
\left(1-\frac{r_-}{r}\right)\,,
\end{equation}
and then, $f(r)$ can also be written as 
\begin{equation}
f(r)=\Delta(r)\sigma^{-2}(r)\,.
\end{equation}

The solution for the electric and dilaton fields are given by
\begin{equation}
F_{01}=-\frac{Q}{r^2}\quad\mbox{and}\quad
\Phi=\Phi_0(r)\equiv\frac{\alpha}{1+\alpha^2}\ln\left(1-\frac{r_-}{r}\right)
=\frac{1}{\alpha}\ln\sigma(r)\,,
\end{equation}
that is,
\begin{equation}
e^{2\alpha\Phi_0(r)}=\sigma^2(r)\,.
\end{equation}

The mass and electric charge of the black hole are given by the formulas:
\cite{GM,GHS,HW,Rakhmanov}
\begin{equation}
M=\frac{1}{2}\left(r_++\frac{1-\alpha^2}{1+\alpha^2}r_-\right)\quad\mbox{and}\quad
Q=\sqrt{\frac{r_+r_-}{1+\alpha^2}}\,,
\label{MQ}
\end{equation}
where we assume $Q>0$ without loss of generality.
Especially, the relation $(1+\alpha^2)Q^2=r_+r_-$ will be used frequently later.

In the above equations, $r_+$ and $r_-$ are called the radii of the outer and
inner horizons \cite{HW}. Strictly speaking, $r=r_-$ does not describe a horizon,
since the dilaton field diverges here. Thus, the limit of $r_-=r_+$ does not
yield an extreme `black hole', but a singularity. We can, however, deal with
such a limit similarly to the point-mass limit of compact objects.
In the limit of $r_-=r_+$, dubbed as the limit of the maximally-charged black 
hole in other words, the mass and charge take the values
\begin{equation}
M=\frac{r_+}{1+\alpha^2}\quad\mbox{and}\quad
Q=\frac{r_+}{\sqrt{1+\alpha^2}}\qquad(r_-=r_+)\,,
\end{equation}
and we find that $r_+>r_-$ means $Q/M<\sqrt{1+\alpha^2}$.

\section{equations with field perturbations}
\label{sec3}

Now, we consider a point mass on the $z$-axis ($\theta=0$), i.e.,
\begin{equation}
T^p_{00}\propto m\delta(\cos\theta-1)\delta(r-b)\,,\quad
J^0_p\propto q\delta(\cos\theta-1)\delta(r-b)\,,\quad
\Sigma^p\propto \beta m\delta(\cos\theta-1)\delta(r-b)\,,
\end{equation}
in the field of the
charged dilatonic black hole discussed in the previous section.

Applying an appropriate gauge \cite{BGR2,RW,Zerilli}, we write down the perturbed
metric as
\begin{eqnarray}
ds^2&=&-[1-\sm H_0(r)Y(\theta)]f(r)dt^2+2\sm H_1(r)Y(\theta)dtdr
+[1+\sm H_2(r)Y(\theta)]f(r)^{-1}dr^2\nonumber
\\
& &+[1+\sm K(r)Y(\theta)]r^2\sigma^2(r)(d\theta^2+\sin^2\theta d\varphi^2)\,,
\label{pmetric}
\end{eqnarray}
where $Y(\theta)\equiv Y_{l0}(\theta)$. Here $Y_{lm}(\theta,\varphi)$ is the
spherical harmonic function. Note that
\begin{equation}
Y_{l0}(\theta)=\sqrt{\frac{2l+1}{4\pi}}P_l(\cos\theta)\,,
\end{equation}
where $P_l(x)$ is the Legendre polynomial.
The labels $l$ is suppressed for variables
(such as $\sm H_{0l}Y_{l0}\rightarrow\sm H_0Y$), as the traditional convention
\cite{BGR2,RW}.

Additionally, we assume that the electromagnetic field takes the form
\begin{equation}
F=\frac{1}{2}F_{\mu\nu}dx^\mu\wedge dx^\nu=\left(-\frac{Q}{r^2}+\sm
f_{01}(r)Y(\theta)\right)dt\wedge dr+\sm f_{02}(r)Y'(\theta)dt\wedge d\theta\,,
\label{pF}
\end{equation}
where $Y'(\theta)\equiv \partial_\theta Y(\theta)$.
Then, the integrability condition $dF=0$ leads to
\begin{equation}
f_{01}-f_{02}'=0\,,
\label{dF}
\end{equation}
where $f_{02}'$ denotes $\partial_r f_{02}$ (and so on).
Besides, we define
\begin{equation}
4\pi J^\mu\equiv\nabla_\nu(e^{-2\alpha\Phi}F^{\mu\nu})\,,
\label{36}
\end{equation}
for later use. With the ansatze so far, one finds that $J^1=J^2=J^3=0$ holds
identically.

In the same way, the dilaton field is assumed to be
\begin{equation}
\Phi=\Phi_0(r)+\sm
\phi(r)Y(\theta)\,.
\label{pPhi}
\end{equation}

The Einstein equation with the point source is rewritten as
\begin{equation}
G_{\mu\nu}-8\pi T^F_{\mu\nu}-8\pi T^\Phi_{\mu\nu}\equiv
\mathcal{G}_{\mu\nu}=8\pi T^p_{\mu\nu}\,,
\end{equation}
where $T^F_{\mu\nu}$ and $T^{\Phi}_{\mu\nu}$ are (\ref{TF}) and (\ref{TPhi})
where the perturbed fields (\ref{pmetric}), (\ref{pF}), and (\ref{pPhi}) are used.
One finds that the
equalities $\mathcal{G}_{03}=\mathcal{G}_{13}=\mathcal{G}_{23}=0$ hold
identically if the perturbed fields (\ref{pmetric}), (\ref{pF}), and (\ref{pPhi})
are adopted.

Since $T^p_{01}=T^p_{02}=0$, the components of the Einstein equation
$\mathcal{G}_{01}$ and $\mathcal{G}_{02}$ should vanish. These equations yields
$H_1(r)=0$, just as in the case with a Reissner--Nordstr\"om black hole
\cite{BGR2}.

We now proceed to calculate the other components of $\mathcal{G}_{\mu\nu}$.
First, we obtain
\begin{eqnarray}
\mathcal{G}_{22}&=&\frac{1}{2}\sm\left[(r-r_+)(r-r_-)
\left(K''(r)-H_0''(r)-\frac{f'(r)}{f(r)}H_0'(r)
+4\Phi_0'(r)\phi'(r)\right)\right.\nonumber \\
& &\quad\left.+
(2r-r_+-r_-)\left(K'(r)-\frac{H_0'(r)+H_2'(r)}{2}\right)
\right.\nonumber \\
& &\left.\quad+l(l+1)(H_0(r)-H_2(r))-\frac{2Q^2}{r^2}(H_0(r)-2\alpha\phi(r))+4Q
f_{01}(r)\right]Y(\theta)\nonumber \\
& &+\frac{1}{2}\sm(H_0(r)-H_2(r))Y''(\theta)\,,
\label{G22}
\end{eqnarray}
where
\begin{equation}
\frac{f'(r)}{f(r)}=
\frac{[(1+\alpha^2)r_++(1-\alpha^2)r_-]r-2r_+r_-}{(1+\alpha^2)r(r-r_+)(r-r_-)}
\quad\mbox{and}\quad
\Phi_0'(r)=
\frac{\alpha}{1+\alpha^2}\frac{r_-}{r(r-r_-)}\,.
\end{equation}

For $l\ge 2$, the Einstein equation $\mathcal{G}_{22}=0$ leads to
$H_0(r)=H_2(r)\equiv W(r)$. Although the cases with
$l=0,1$ should be considered separately \cite{BGR2}, we take $H_0=H_2=W$ as an
ansatz. Now, the equation (\ref{G22}) reads
\begin{eqnarray}
& &(r-r_+)(r-r_-)\left(K''-W''-\frac{f'}{f}W'+4\Phi_0'\phi'\right)+(2r-r_+-r_-)
\left(K'-W'\right)\nonumber \\ &
&\quad-\frac{2Q^2}{r^2}(W-2\alpha\phi)+4Qf_{01}=0\,.
\label{G22b}
\end{eqnarray}
On the other hand, 
from the equation $\mathcal{G}_{11}=0$, one obtains
\begin{eqnarray}
& &f_{01}=\frac{1}{4Q}\Bigl[(2r-r_+-r_-)(K'-W')+(r-r_+)(r-r_-)
\left(\frac{f'}{f}W'-
4\Phi_0'\phi'\right)\nonumber
\\ & &\qquad\qquad-[l(l+1)-2](K-W)+\frac{2Q^2}{r^2}(W-2\alpha\phi)\Bigr]\,.
\label{f01}
\end{eqnarray}
Substitute (\ref{f01}) into (\ref{G22b}), we obtain
\begin{equation}
(r-r_+)(r-r_-)(K-W)''+2(2r-r_+-r_-)(K-W)'-[l(l+1)-2](K-W)=0\,.
\label{eqX}
\end{equation}
Note that this equation does not include the modes of the dilaton field $\phi$.
From the equation $\mathcal{G}_{12}=0$, one obtains
\begin{equation}
f_{02}=-\frac{r^2\Delta(r)}{4Q}\left[K'-W'-\frac{f'}{f}W+
4\Phi_0'\phi\right]\,.
\label{f02}
\end{equation}
Substituting (\ref{f01}) and (\ref{f02}) into the integrability equation
(\ref{dF}), we obtain the same equation (\ref{eqX}).

The only regular solution in the region $[r_+,\infty]$ for
$(r-r_+)(r-r_-)X''+2(2r-r_+-r_-)X'-[l(l+1)-2]X=0$ is $X(r)\equiv 0$ \cite{BGR2},
so we should take $K(r)=W(r)$.
At this stage, the linear-perturbation modes of the electric field strength are
expressed by
\begin{eqnarray}
f_{01}&=&\frac{1}{4Q}\left[r^2\Delta(r)
\left(\frac{f'}{f}W'-
4\Phi_0'\phi'\right)
+\frac{2Q^2}{r^2}(W-2\alpha\phi)\right]\,,
\label{f01r}
\\
f_{02}&=&\frac{r^2\Delta(r)}{4Q}\left[\frac{f'}{f}W-
4\Phi_0'\phi\right]\,,
\label{f02r}
\end{eqnarray}
which will be used later. Note that the condition (\ref{dF}) can be confirmed
by use of (\ref{f01r}) and (\ref{f02r}), without using the equation of motion for
$W$ and $\phi$.

Now, setting $K=W$, we find that $\mathcal{G}_{00}$ with (\ref{f01})
takes the form
\begin{equation}
\mathcal{G}_{00}=-f^2(r)
\sm\left[W''(r)+\frac{2}{r}W'(r)
-\frac{l(l+1)}{r^2\Delta(r)}W(r)+2\Phi_0'(r)(\alpha
W'(r)+2\phi'(r))
\right]Y(\theta)\,,
\end{equation}
while $4\pi J^0$ (\ref{36}) can be written, with all the information thus far, as
\begin{eqnarray}
4\pi J^0&=&-\frac{r^2f(r)}{4Q}\sm\left[\frac{f'(r)}{f(r)}
\left(W''(r)+\frac{2}{r}W'(r)-\frac{l(l+1)}{r^2\Delta(r)}W(r)\right)\right.
\nonumber \\
& &\qquad\qquad\qquad\left.-4\Phi_0'(r)
\left(\phi''(r)+\frac{2}{r}\phi'(r)-\frac{l(l+1)}{r^2\Delta(r)}\phi(r)\right)
\right]Y(\theta)\,.
\end{eqnarray}
In addition, the following quantity including the second-order derivative
of the dilaton field becomes
\begin{eqnarray}
4\pi\Sigma&\equiv&\nabla^2\Phi+\frac{\alpha}{2}e^{-2\alpha\Phi}F^2\nonumber
\\ &=&f(r)\sm\left[\phi''(r)+\frac{2}{r}\phi'(r)
-\frac{l(l+1)}{r^2\Delta(r)}\phi(r)
+\frac{f'(r)}{2f(r)}(\alpha W'(r)+2\phi'(r))\right]
Y(\theta)\,.
\end{eqnarray}

Now, leaving them aside, let us fix the rigorous form of the point source.
The action of a particle coupled to dilaton is given by \cite{MS}
\begin{equation}
S_p=-\int \left[me^{\beta\Phi}+qA_\mu\frac{d x^\mu}{d\tau}\right]d\tau\,.
\end{equation}
Presuming the action, we decipher the form of the sources of gravitational,
electrical, and dilaton fields as
\begin{eqnarray}
T_p^{\mu\nu}&=&\frac{m}{\sqrt{-g}}
\int e^{\beta\Phi(x)}\delta^4(x-z(\tau))U^\mu U^\nu d\tau\,,
\\
J^0_p&=&\frac{q}{\sqrt{-g}}
\int \delta^4(x-z(\tau))U^\mu d\tau\,,
\\
\Sigma^p&=&\frac{\beta m}{\sqrt{-g}}
\int e^{\beta\Phi(x)}\delta^4(x-z(\tau)) d\tau\,,
\end{eqnarray}
where $U^\mu\equiv\frac{dz^\mu}{d\tau}$ and  the normalization of the delta
function is defined by $\int\delta^4(x)d^4x=1$.
Further, for the static point located at $r=b$ on the $z$-axis $(\theta=0)$
in the background of the charged dilaton black hole, we find%
\footnote{Note that, since the particle is located  at $\theta=0$, the factor
$2\pi$ appears instead of $\delta(\varphi-\varphi_0)$.}
\begin{eqnarray}
T^p_{00}&=&\frac{me^{\beta\Phi_0(b)}}{2\pi
b^2\sigma^2(b)}f(b)^{3/2}\delta(r-b)\delta(\cos\theta-1)\,,
\\
J^0_p&=&\frac{q}{2\pi b^2\sigma^2(b)}\delta(r-b)\delta(\cos\theta-1)\,,
\\
\Sigma^p&=&\frac{\beta me^{\beta\Phi_0(b)}}{2\pi
b^2\sigma^2(b)}f(b)^{1/2}\delta(r-b)\delta(\cos\theta-1)\,,
\end{eqnarray}
since $U^0=\frac{1}{\sqrt{f}}$ and $d\tau=\sqrt{f}dt$, in this case \cite{BGR2}.

For now, the remaining equations are
\begin{equation}
\mathcal{G}_{00}=8\pi T^p_{00}\,,\quad
4\pi J^0=4\pi J_p^0\,,\quad
4\pi \Sigma=4\pi \Sigma^p\,.
\label{remeq}
\end{equation}
We now have three coupled equations for two functions $W$ and $\phi$ to be
solved.  The compatibility of the system of equations requires 
\begin{equation}
me^{\beta\Phi_0(b)}\left(\frac{f'(b)}{f(b)}+2\beta\Phi_0'\right)
=\frac{2qQ}{b^2\sqrt{f(b)}}\,.
\label{static}
\end{equation}
This relation is the same as the static condition of a test particle in the
background of a charged dilaton black hole \cite{MS}. Namely, the equation
$V_{eff}'(b)=0$, where
\begin{equation}
V_{eff}(r)=\frac{qQ}{r}+m\sqrt{f(r)}e^{\beta\Phi_0(r)}\,,
\end{equation}
is the effective potential for the particle (with vanishing angular momentum
\cite{MS}).%
\footnote{At large distance,
$V_{eff}(r)=\frac{qQ-mM-\Sigma_p(\infty)\Sigma_{bh}(\infty)}{r}+O(r^{-2})$, where
$\Sigma_p(\infty)=\beta m$ and $\Sigma_{bh}(\infty)=\frac{\alpha
r_-}{1+\alpha^2}$ (see Sec.~\ref{sec5}).}

\section{the case with $\beta=\alpha$}
\label{sec4}
The differential equations (\ref{remeq}) are clearly simplified if
$\phi(r)=-\frac{\alpha}{2}W(r)$.
Then, the equations become
\begin{eqnarray}
8\pi T^p_{00}&=&-f^2(r)
\sm\left[W''(r)+\frac{2}{r}W'(r)
-\frac{l(l+1)}{r^2\Delta(r)}W(r)
\right]Y(\theta)\,,
\\
4\pi J_p^0&=&-\frac{r^2f(r)}{4Q}\frac{\Delta'(r)}{\Delta(r)}\sm\left[
W''(r)+\frac{2}{r}W'(r)-\frac{l(l+1)}{r^2\Delta(r)}W(r)
\right]Y(\theta)\,,
\\
4\pi\Sigma^p&=&-\frac{\alpha}{2}f(r)\sm\left[W''(r)+\frac{2}{r}W'(r)
-\frac{l(l+1)}{r^2\Delta(r)}W(r)\right]
Y(\theta)\,,
\end{eqnarray}
and then, the equations are compatible if and only if
\begin{equation}
\beta=\alpha\quad\mbox{and}\quad
m\frac{\Delta'(b)}{2\sqrt{\Delta(b)}}=\frac{qQ}{b^2}\,.
\label{ccond}
\end{equation}
This stability condition reads
\begin{equation}
\frac{q}{m\sqrt{1+\alpha^2}}=\frac{b^2\Delta'(b)}{2\sqrt{r_+r_-\Delta(b)}}\,.
\end{equation}
Since $(b^2\Delta'(b))^2-(2\sqrt{r_+r_-\Delta(b)})^2=(r_+-r_-)^2\ge 0$,
the charge-to-mass ratio must satisfy the condition
$q/m\ge\sqrt{1+\alpha^2}$.
Especially, the equality $q/m=\sqrt{1+\alpha^2}$ is achieved if and only if
the black hole is maximally charged (that is, $Q/M=\sqrt{1+\alpha^2}$), and
the equilibrium realizes for any value of $b$ then.

In this case, the sources are written as
\begin{eqnarray}
T^p_{00}&=&\frac{m}{2\pi
b^2\sigma(b)}f(b)^{3/2}\delta(r-b)\delta(\cos\theta-1)\,,
\\
J^0_p&=&\frac{q}{2\pi b^2\sigma^2(b)}\delta(r-b)\delta(\cos\theta-1)\,,
\\
\Sigma^p&=&\frac{\alpha m}{2\pi
b^2\sigma(b)}f(b)^{1/2}\delta(r-b)\delta(\cos\theta-1)\,,
\end{eqnarray}
and we obtain the following single partial differential equation:
\begin{equation}
\left[\partial_r^2+\frac{2}{r}\partial_r
+\frac{1}{r^2\Delta(r)}
\left(\partial_\theta^2+\cot\theta\partial_\theta\right)
\right]\mathcal{W}(r,\theta)=-\frac{m}{2\pi
b^2\sqrt{\Delta(b)}}\delta(r-b)\delta(\cos\theta-1)\,,
\end{equation}
where $\mathcal{W}(r,\theta)=\sm W(r)Y(\theta)$.

As can be seen from the form of the equation,
the regular solution of this equation \cite{LL} is exactly the same as for the
Reissner--Nordstr\"om black hole \cite{BGR1,BGR2,BGR3,BGR4}, if it is expressed
in terms of two radii $r_+$ and $r_-$. That is,
\begin{eqnarray}
\mathcal{W}(r,\theta)&=&
\frac{2m}{
b\sqrt{\Delta(b)}}\frac{1}{r\mathcal{D}}\left[
\left(r-\pl\right)\left(b-\pl\right)
-\left(\mi\right)^2\cos\theta\right]\nonumber \\
&=&
\frac{4Qq}{
b^3{\Delta'(b)}}\frac{1}{r\mathcal{D}}\left[
\left(r-\pl\right)\left(b-\pl\right)
-\left(\mi\right)^2\cos\theta\right]\,,
\end{eqnarray}
where
\begin{eqnarray}
\mathcal{D}(r,\theta)&=&\left[\left(r-\pl\right)^2+
\left(b-\pl\right)^2\right.\nonumber \\
& &\left.-2
\left(r-\pl\right)\left(b-\pl\right)\cos\theta
-\left(\mi\right)^2\sin^2\theta\right]^{1/2}\,.
\end{eqnarray}
Note that the relation
\begin{equation}
\partial_\theta\mathcal{D}=\frac{\sin\theta}{\mathcal{D}}\left[
\left(r-\pl\right)\left(b-\pl\right)
-\left(\mi\right)^2\cos\theta\right]\,
\end{equation}
will be used later.

In this case, the dilaton field is given by
\begin{equation}
\Phi=\Phi_0(r)+\sm\phi(r)Y(\theta)=\Phi_0(r)+\tilde{\phi}(r,\theta)
=\Phi_0(r)-\frac{\alpha}{2}\mathcal{W}(r,\theta)\,.
\end{equation}

Besides the coupled differential equations can be simply solved in this case, 
the condition $\beta=\alpha$ is very interesting because the point source can be
regarded as a singular limit of the charged dilaton black hole in the system
governed by the single action (\ref{ac}), although the massive point is a limit
of an over-charged naked singularity (i.e., $q/m>\sqrt{1+\alpha^2}$) except for
the maximally-charged dilaton black hole (i.e., $r_-=r_+$).

\section{Electric and dilatonic charges and mass of the system 
($\beta=\alpha$)}
\label{sec5}

\subsection{the electric field and charge ($\beta=\alpha$)}
In the case with $\beta=\alpha$, which means $\phi=-\frac{\alpha}{2}W$, the
linear-perturbation modes of the electric field strength (\ref{f01r}) and
(\ref{f02r}) take simpler forms:
\begin{equation}
f_{01}=\frac{1}{4Q}\left[r^2
{\Delta'}W'
+\frac{2(1+\alpha^2)Q^2}{r^2}W\right]
=\frac{1}{4Q}\left[r^2
{\Delta'}W'
+\frac{2r_+r_-}{r^2}W\right]\,,
\quad
f_{02}=\frac{r^2{\Delta'}}{4Q}W\,.
\end{equation}

Thus, the electrostatic potential $V_p=-A_0$ for the point-charge contribution
is given by
\begin{equation}
V_p(r,\theta)=\frac{r^2\Delta'(r)}{4Q}\mathcal{W}(r,\theta)
=\frac{qr{\Delta'(r)}}{
b^3{\Delta'(b)}}\frac{1}{\mathcal{D}}\left[
\left(r-\pl\right)\left(b-\pl\right)
-\left(\mi\right)^2\cos\theta\right]\,,
\end{equation}
which leads to
\begin{equation}
F=-\left(\frac{Q}{r^2}+E_r\right)dt\wedge dr-E_\theta dt\wedge
d\theta\,,
\end{equation}
where
\begin{equation}
E_r=-\sm f_{01}(r)Y(\theta)=-\partial_rV_p(r,\theta)\,,\quad
E_\theta=-\sm f_{02}(r)Y'(\theta)=-\partial_\theta V_p(r,\theta)
\end{equation}

To the first order in the perturbation, the electrical flux $4\pi Q_{bh}(r)$
generated by the charged dilaton black hole is given by
\begin{eqnarray}
4\pi Q_{bh}(r)&=&2\pi\int_0^\pi
r^2\sigma^2e^{-2\alpha\Phi}(1+\mathcal{W})\frac{Q}{r^2}\sin\theta
d\theta=2\pi\int_0^\pi
r^2\sigma^2e^{-2\alpha\Phi_0}(1+\mathcal{W}+\alpha^2\mathcal{W})\frac{Q}{r^2}
\sin\theta d\theta
\nonumber \\
&=&4\pi Q+2\pi
\frac{4(1+\alpha^2)Q^2q}{
rb^3{\Delta'(b)}}\int_0^\pi\frac{\sin\theta}{\mathcal{D}}\left[
\left(r-\pl\right)\left(b-\pl\right)
-\left(\mi\right)^2\cos\theta\right]d\theta\nonumber \\
&=&4\pi Q+\frac{8\pi qr_+r_-}{
rb^3{\Delta'(b)}}[\mathcal{D}(r,\pi)-\mathcal{D}(r,0)]\,,
\end{eqnarray}
while the electrical flux $4\pi Q_p(r)$ generated by the point charge is given by
\begin{eqnarray}
4\pi Q_p(r)&=&2\pi\int_0^\pi r^2\sigma^2e^{-2\alpha\Phi_0}E_r \sin\theta d\theta
\nonumber \\
&=&-2\pi r^2\partial_r\left\{
\frac{qr{\Delta'(r)}}{
b^3{\Delta'(b)}}\int_0^\pi\frac{\sin\theta}{\mathcal{D}}\left[
\left(r-\pl\right)\left(b-\pl\right)
-\left(\mi\right)^2\cos\theta\right]d\theta\right\}
\nonumber \\
&=&-2\pi r^2\partial_r\left\{
\frac{qr{\Delta'(r)}}{
b^3{\Delta'(b)}}[\mathcal{D}(r,\pi)-\mathcal{D}(r,0)]\right\}
\nonumber \\
&=&2\pi \frac{q}{b^3{\Delta'(b)}}\Bigl\{
\left(r_++r_--\frac{4r_+r_-}{r}\right)[\mathcal{D}(r,\pi)-\mathcal{D}(r,0)]
\nonumber \\
& &\qquad\qquad\qquad-
{r^3{\Delta'(r)}}\partial_r[\mathcal{D}(r,\pi)-\mathcal{D}(r,0)]\Bigr\}\,.
\end{eqnarray}

Thus, the total flux $4\pi Q_{tot}(r)\equiv 4\pi Q_{bh}(r)+4\pi Q_{p}(r)$ turns
out to be
\begin{eqnarray}
4\pi Q_{tot}(r)&=&4\pi Q+2\pi \frac{q}{b^3{\Delta'(b)}}\Bigl\{
\left(r_++r_-\right)[\mathcal{D}(r,\pi)-\mathcal{D}(r,0)]-
{r^3{\Delta'(r)}}\partial_r[\mathcal{D}(r,\pi)-\mathcal{D}(r,0)]\Bigr\}
\nonumber \\
&=&4\pi Q+4\pi \frac{q}{b^3{\Delta'(b)}}\Bigl\{
\left(r_++r_-\right)[(r-\pl)\vartheta(b-r)+(b-\pl)\vartheta(r-b)]
\nonumber \\
& &\qquad\qquad\qquad\qquad-
[(r_++r_-)r-2r_+r_-]\vartheta(b-r)\Bigr\}
\nonumber \\
&=&4\pi Q+4\pi \frac{q}{b^3{\Delta'(b)}}\Bigl\{
-2\left(\mi\right)^2\vartheta(b-r)+(r_++r_-)(b-\pl)\vartheta(r-b)\Bigr\}
\nonumber \\
&=&4\pi Q+4\pi q\vartheta(r-b)-4\pi \frac{2q}{b^3{\Delta'(b)}}
\left(\mi\right)^2\,,
\end{eqnarray}
where the function $\vartheta(x)$ denotes a step function, i.e.,
$\vartheta(x)=1$ for $x>0$ and $\vartheta(x)=0$ for $x<0$.

Therefore, we obtain
\begin{equation}
4\pi Q_{bh}(r)+4\pi Q_p(r)+4\pi \bar{Q}=4\pi Q+4\pi q\vartheta(r-b)\,,
\end{equation}
where
\begin{equation}
4\pi \bar{Q}\equiv 4\pi \frac{2q}{b^3{\Delta'(b)}}
\left(\frac{r_+-r_-}{2}\right)^2\,.
\end{equation}
The charge $\bar{Q}$ can be recognized by the induced charge at the horizon,
or this can be regarded as that from the $l=0$ mode of $f_{01}$, i.e., an
additional contribution $\sim\bar{Q}/r^2$ to $f_{01}$, which has been omitted in
the analysis so far. This mode corresponds to the constant shift ($l=0$) in
$\mathcal{W}$ and $\tilde{\phi}$, which can be eliminated by a suitable
transformation \cite{BGR2}.%
\footnote{The treatment by replacing $Q\rightarrow Q-\bar{Q}$ is suggested in
the references \cite{BGR3,BGR4}.}

Note that $4\pi \bar{Q}=0$ automatically if and only if $r_-=r_+$, that is the
case with the maximally-charged dilaton black hole.
In the limit of $r_-=r_+$, the compatibility condition (\ref{ccond}) becomes
\begin{equation}
mr_+=qQ\qquad(r_-=r_+)\,,
\end{equation}
which is independent of the location of the point charge, $b$.
The electrostatic potential $V_p$ in the limit of $r_-=r_+$ reads
\begin{equation}
\left.V_p(r,\theta)\right|_{r_-=r_+}=
\frac{q\left(r-r_+\right)^2}{r^2\sqrt{\left(r-r_+\right)^2+
\left(b-r_+\right)^2-2
\left(r-r_+\right)\left(b-r_+\right)\cos\theta}}\,.
\label{Vp}
\end{equation}
\begin{figure}[ht]
\centering
\includegraphics[width=6cm]
{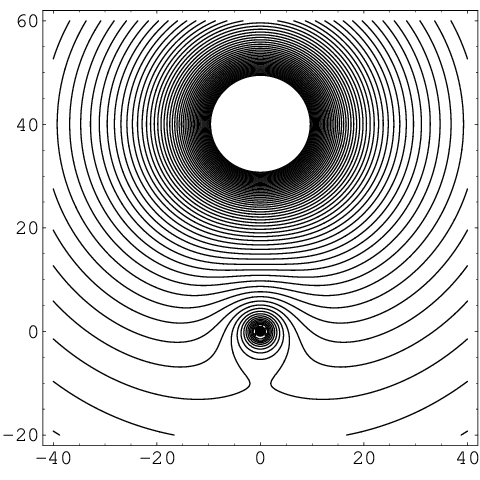}
\caption{The contour plot of the electrostatic potential $V_p$ of the
particle in the $x$-$z$ plane ($x=r\sin\theta$ and $z=r\cos\theta$ are
Cartesian-like coordinates) for $r_-=r_+=1$ and $b=40$.}
\label{fig1}
\end{figure}
In FIG.~\ref{fig1}, we show the contour plot of the electrostatic potential $V_p$
for $r_-=r_+=1$ and $b=40$.\footnote{Here, $q$ or the height of each contour line
is considered arbitrary.}
Thus, the electric field $E_r=-\partial_r V_p$ produced by 
the point charge vanishes at the horizon, $r=r_+$.
This is just the electric Meissner effect found in \cite{BGR1,BGR2,BGR3,BGR4},
because the electric line of force originated from the point charge cannot cross
the horizon. 
Here, we would like to emphasize that if $V_p$ for $r_-=r_+$ is expressed by $r_+$
and $b$, it is completely the same as the case of the Reissner--Nordstr\"om
black hole, while the value of $r_+$ and the condition for $r_-=r_+$ expressed by
mass, charge, and dilaton coupling constant are different.

\subsection{the dilatonic field and charge ($\beta=\alpha$)}
The dilatonic flux $4\pi \Sigma_{bh}$ of the charged dilaton black hole alone
cannot be defined on an arbitrary closed surface around the black hole,
because the dilaton is coupled to the Maxwell field strength in the present
model and the dilaton charge is not the Noether charge, contrarily to the electric
charge. The scalar charge can simply be defined at spatial infinity in several
cases \cite{GHS,Rakhmanov,GKK,Pacilio,BGOZ}.

We consider the following integration on the sphere $S^2$ centered at the origin
with the radius
$r$ as the dilatonic flux:
\begin{equation}
4\pi\Sigma(r)\equiv\int_{S^2}\sqrt{-g}g^{rr}\partial_r\Phi dS\,,
\end{equation}
that is, the asymptotic behavior of the dilaton field is
$\Phi\sim-\frac{\Sigma(\infty)}{r}$ ($r\rightarrow\infty$).

The contribution solely from the black hole can be found as
\begin{equation}
4\pi\Sigma_{bh}(r)\equiv 4\pi
r^2\sigma^2(r)f(r)\Phi_0'(r)=4\pi
r^2\Delta(r)\Phi_0'(r)
=\frac{4\pi\alpha r_-}{1+\alpha^2}
\left(1-\frac{r_+}{r}\right)\,.
\end{equation}
Note that $\Sigma_{bh}(\infty)=\frac{\alpha r_-}{1+\alpha^2}$
and $\Sigma_{bh}(r_+)=0$.

Since the perturbative dilaton field which comes from the point source is
\begin{equation}
\tilde{\phi}(r,\theta)=-\frac{\alpha m}{
b\sqrt{\Delta(b)}}\frac{1}{r\mathcal{D}}\left[
\left(r-\pl\right)\left(b-\pl\right)
-\left(\mi\right)^2\cos\theta\right]\,,
\end{equation}
the dilatonic flux generated by the point charge is given by
\begin{eqnarray}
4\pi\Sigma_p(r)&=&2\pi\int_0^\pi r^2\Delta(r) \partial_r\tilde{\phi}(r,\theta)
\sin\theta
d\theta
\nonumber \\
&=&-2\pi\frac{\alpha m\Delta(r)}{b\sqrt{\Delta(b)}} r^2\partial_r\left\{
\frac{1}{r}\int_0^\pi\frac{\sin\theta}{\mathcal{D}}\left[
\left(r-\pl\right)\left(b-\pl\right)
-\left(\mi\right)^2\cos\theta\right]d\theta\right\}
\nonumber \\
&=&-2\pi\frac{\alpha m\Delta(r)}{b\sqrt{\Delta(b)}} r^2\partial_r\left\{
\frac{1}{r}[\mathcal{D}(r,\pi)-\mathcal{D}(r,0)]\right\}
\nonumber \\
&=&2\pi \frac{\alpha m\Delta(r)}{b\sqrt{\Delta(b)}}\Bigl\{
[\mathcal{D}(r,\pi)-\mathcal{D}(r,0)]-
r\partial_r[\mathcal{D}(r,\pi)-\mathcal{D}(r,0)]\Bigr\}
\nonumber \\
&=&4\pi \frac{\alpha m\Delta(r)}{b\sqrt{\Delta(b)}}\Bigl\{
[(r-\pl)\vartheta(b-r)+(b-\pl)\vartheta(r-b)]-r\vartheta(b-r)\Bigr\}\nonumber \\
&=&4\pi \frac{\alpha m\Delta(r)}{b\sqrt{\Delta(b)}}\Bigl[
-\pl+b\vartheta(r-b)\Bigr]\,.
\end{eqnarray}
It is notable that the discontinuity of $\Sigma_p(r)$ appears at $r=b$.
Note that $\Sigma_p(r_+)=0$ and
\begin{equation}
\Sigma_p(\infty)=\frac{\alpha m}{b\sqrt{\Delta(b)}}\Bigl[
b-\pl\Bigr]\,.
\end{equation}
It is notable that, in the limit of $b\rightarrow\infty$,
$\Sigma_p(\infty)\rightarrow\alpha m$.

Another interesting result is obtained if $r_-=r_+$, that is
\begin{equation}
\Sigma_{bh}(\infty)=\alpha M\,,\quad\Sigma_p(\infty)=\alpha m\qquad(r_-=r_+)
\end{equation}
are obtained, which are independent of the value of $b$.
Because the square of the electric fields $F^2$ is also the source of the
dilatonic filed, it is natural that odd values come out for $\Sigma(\infty)$.
Nevertheless, it is remarkable that the point-particle description has
validity in the limit of the maximal charge of the black hole.

\subsection{mass of the system ($\beta=\alpha$)}

The asymptotic mass observed at the spatial infinity is given by \cite{BGR1,BGR2}
\begin{equation}
M_{tot}=\frac{1}{2}\lim_{r\rightarrow\infty}r\Bigl[1-f(r)(1-\mathcal{W}(r,\theta))
\Bigr]\,.
\end{equation}
Incidentally, this formula simply corresponds to the fact that the time-time
component of the metric $g_{tt}$ can be approximated as
\begin{equation}
g_{tt}=-\left(1-\frac{2M_{tot}}{r}\right)+O(r^{-2})\,,
\end{equation}
in the asymptotic region.
  Then, for the case with $\beta=\alpha$, we
obtain
\begin{equation}
M_{tot}=M+\frac{m}{\sqrt{\Delta(b)}}\left(1-\frac{r_++r_-}{2b}\right)\,,
\label{mtot}
\end{equation}
noting that up to the first order of $m$ is meaningful. Here, $M$ is given by
(\ref{MQ}).

We can rewrite (\ref{mtot}) and find that
\begin{equation}
M_{tot}=M+m\frac{(b-r_+)+(b-r_-)}{2\sqrt{(b-r_+)(b-r_-)}}\ge M+m\,,
\end{equation}
where we used the inequality of arithmetic and geometric means,
i.e., $(A+B)/2\ge\sqrt{AB}$ for positive $A$ and $B$.
The equality holds if $r_-=r_+$.

\section{comparison with the exact solution of the maximally charged black holes
($\beta=\alpha$)}
\label{sec6}

If we suppose $\beta=\alpha$ and $r_-=r_+$, which is the limit of the
maximal charge of the charged dilaton black hole, the equation for the equilibrium
condition
$V'(b)=0$ becomes independent of
$b$, 
\begin{equation}
qQ=mr_+=mM(1+\alpha^2)=mQ\sqrt{1+\alpha^2}\,,
\end{equation}
that is,
\begin{equation}
\frac{q}{m}=\frac{Q}{M}=\sqrt{1+\alpha^2}\,.
\end{equation}
Thus, the charge and mass of the point particle should satisfy the same
relation as the charged dilaton black hole if $\beta=\alpha$ and $r_-=r_+$.

Noting that $r_-=r_+$, we obtain
\begin{eqnarray}
\left.\mathcal{W}(r,\theta)\right|_{r_-=r_+}&=&
\frac{2m\left(r-r_+\right)}{r\sqrt{\left(r-r_+\right)^2+
\left(b-r_+\right)^2-2
\left(r-r_+\right)\left(b-r_+\right)\cos\theta}}\,,
\\
\left.V_p(r,\theta)\right|_{r_-=r_+}
&=&\frac{q\left(r-r_+\right)^2}{r^2\sqrt{\left(r-r_+\right)^2+
\left(b-r_+\right)^2-2
\left(r-r_+\right)\left(b-r_+\right)\cos\theta}}\,.
\end{eqnarray}

On the other hand, the exact solution for two maximally charged dilaton black
holes is known as
\cite{Shiraishi}
\begin{equation}
ds^2=-U^{-\frac{2}{1+\alpha^2}}dt^2
+U^{\frac{2}{1+\alpha^2}}
[dR^2+R^2(d\theta^2+\sin^2\theta d\varphi^2)]\,,
\label{ex1}
\end{equation}
and
\begin{equation}
-A_0=\frac{1}{\sqrt{1+\alpha^2}}(1-U^{-1})\,,\quad
e^{-2\alpha\Phi}=U^{\frac{2\alpha^2}{1+\alpha^2}}\,,
\label{ex2}
\end{equation}
where
\begin{equation}
U=1+\frac{r_+}{R}+\frac{(1+\alpha^2)m}{\sqrt{R^2+B^2-2RB\cos\theta}}\,.
\end{equation}

Changing the coordinates as
$R=r-r_+$, $B=b-r_+$, $U$ reads 
\begin{equation}
U=\frac{1}{1-\frac{r_+}{r}}\left[1+
\frac{(1+\alpha^2)m\left(1-\frac{r_+}{r}\right)}%
{\sqrt{(r-r_+)^2+(b-r_+)^2-2(r-r_+)(b-r_+)\cos\theta}}\right]\,.
\end{equation}

The straightforward calculation reveals that (\ref{ex1}) and (\ref{ex2})
can be expressed, in the first order of $m$, as
\begin{eqnarray}
& &ds^2=-\left(1-\frac{r_+}{r}\right)^{\frac{2}{1+\alpha^2}}
\left[1-\left.\mathcal{W}(r,\theta)\right|_{r_-=r_+}\right]dt^2
\nonumber \\
& &+\left[1+\left.\mathcal{W}(r,\theta)\right|_{r_-=r_+}\right]
\left[\left(1-\frac{r_+}{r}\right)^{-\frac{2}{1+\alpha^2}}dr^2+r^2
\left(1-\frac{r_+}{r}\right)^{\frac{2\alpha^2}{1+\alpha^2}}
(d\theta^2+\sin^2\theta
d\varphi^2)\right]\,,
\end{eqnarray}
and
\begin{equation}
-A_0=\frac{Q}{r}+\left.V_p(r,\theta)\right|_{r_-=r_+}\,,\quad
\Phi=\frac{\alpha}{1+\alpha^2}\ln\left(1-\frac{r_+}{r}\right)-\frac{\alpha}{2}
\left.\mathcal{W}(r,\theta)\right|_{r_-=r_+}\,.
\end{equation}
Thus, we confirmed the perturbative calculation for $\beta=\alpha$ and $r_-=r_+$
coincides with the exact solution of two maximally-charged black holes
in the first order of $m=q/\sqrt{1+\alpha^2}$.

\section{the electric field around a maximally charged dilaton black hole
$r_-=r_+$ ($\beta\ne\alpha$)}
\label{sec7}

In Sec.~\ref{sec5}, we have found the electric Meissner effect for the case with
$\beta=\alpha$ in the limit of
$r_-=r_+$. In this section, we investigate the case in
the limit of $r_-=r_+$, but $\beta\ne\alpha$.
First, we notice that, when $r_-=r_+$,
\begin{equation}
f(r)=\left(1-\frac{r_+}{r}\right)^\frac{2}{1+\alpha^2}\,,\quad
\sigma(r)=
\left(1-\frac{r_+}{r}\right)^\frac{\alpha^2}{1+\alpha^2}\,,\quad
\Phi_0(r)=\frac{\alpha}{1+\alpha^2}\ln\left(1-\frac{r_+}{r}\right)\,,
\label{mc}
\end{equation}
and
\begin{equation}
\Delta(r)=\left(1-\frac{r_+}{r}\right)^2\,.
\label{mc2}
\end{equation}
Consequently, we find
\begin{equation}
\frac{f'(r)}{f(r)}=\frac{2}{1+\alpha^2}
\frac{r_+}{r(r-r_+)}\quad\mbox{and}\quad
\Phi_0'(r)=
\frac{\alpha}{1+\alpha^2}\frac{r_+}{r(r-r_+)}\,.
\end{equation}

The two independent differential equations are found to be
\begin{eqnarray}
8\pi T^p_{00}&=&-f^2(r)
\sm\left[W''(r)+\frac{2}{r}W'(r)
-\frac{l(l+1)}{(r-r_+)^2}W(r)\right.\nonumber \\
& &\qquad\qquad\left.+\frac{2\alpha}{1+\alpha^2}\frac{r_+}{r(r-r_+)}(\alpha
W'(r)+2\phi'(r))
\right]Y(\theta)\,,\label{74}
\\
4\pi\Sigma^p&=&f(r)\sm\left[\phi''(r)+\frac{2}{r}\phi'(r)
-\frac{l(l+1)}{(r-r_+)^2}\phi(r)\right.\nonumber \\
& &\left.\qquad\qquad
+\frac{1}{1+\alpha^2}
\frac{r_+}{r(r-r_+)}(\alpha W'(r)+2\phi'(r))\right]
Y(\theta)\,,\label{75}
\end{eqnarray}
where
\begin{eqnarray}
T^p_{00}&=&\frac{m}{2\pi
b^2\sigma^2(b)}\left(1-\frac{r_+}{b}\right)^{\frac{\alpha\beta}{1+\alpha^2}}f(b)^{3/2}\delta(r-b)\delta(\cos\theta-1)\,,
\label{76}\\
\Sigma^p&=&\frac{\beta m}{2\pi
b^2\sigma^2(b)}
\left(1-\frac{r_+}{b}\right)^{\frac{\alpha\beta}{1+\alpha^2}}f(b)^{1/2}
\delta(r-b)\delta(\cos\theta-1)\,.\label{77}
\end{eqnarray}
The compatibility (equilibrium) condition (\ref{static}) for $r_-=r_+$ becomes
\begin{equation}
\frac{q}{m\sqrt{1+\alpha^2}}=\frac{1+\alpha\beta}{1+\alpha^2}
\left(1-\frac{r_+}{b}\right)^{\frac{\alpha(\beta-\alpha)}{1+\alpha^2}}\,.
\label{78}
\end{equation}

The linear combination of (\ref{76}) and (\ref{77}) with (\ref{78})
yields the partial differential equation
\begin{equation}
\left[\partial_r^2+\frac{2}{r}\partial_r
+\frac{1}{(r-r_+)^2}
\left(\partial_\theta^2+\cot\theta\partial_\theta\right)
\right](\mathcal{W}-2\alpha\tilde{\phi})=-\frac{\sqrt{1+\alpha^2}q}{2\pi
b^2\sqrt{\Delta(b)}}\delta(r-b)\delta(\cos\theta-1)\,.
\end{equation}
The regular solution of the equation, which we have already known, is expressed as
\begin{equation}
\mathcal{W}(r,\theta)-2\alpha\tilde{\phi}(r,\theta)=
\frac{2\sqrt{1+\alpha^2}q\left(r-r_+\right)}{r\sqrt{\left(r-r_+\right)^2+
\left(b-r_+\right)^2-2
\left(r-r_+\right)\left(b-r_+\right)\cos\theta}}\,.
\end{equation}

Remarkably, using (\ref{mc}), (\ref{mc2}), and $\sqrt{1+\alpha^2}Q=r_+$, the modes
of the electric field strength (\ref{f01r}) and (\ref{f02r}) become
\begin{eqnarray}
f_{01}&=&\frac{1}{2\sqrt{1+\alpha^2}}\left[
\left(1-\frac{r_+}{r}\right)(W'-2\alpha\phi')
+\frac{r_+}{r^2}(W-2\alpha\phi)\right]\qquad (r_-=r_+)\,,
\\
f_{02}&=&\frac{1}{2\sqrt{1+\alpha^2}}
\left(1-\frac{r_+}{r}\right)(W-2\alpha\phi)\qquad\qquad\qquad\qquad
\qquad\quad (r_-=r_+)\,.
\end{eqnarray}
Therefore, we obtain the following electrostatic potential:
\begin{equation}
\left.V_p(r,\theta)\right|_{r_-=r_+}
=\frac{q\left(r-r_+\right)^2}{r^2\sqrt{\left(r-r_+\right)^2+
\left(b-r_+\right)^2-2
\left(r-r_+\right)\left(b-r_+\right)\cos\theta}}\,.
\end{equation}
Note that this is identical to (\ref{Vp}).
Apparently, the radial component of the electric field $E_r=-\partial_rV_p$
originated from the point charge vanishes at the horizon, $r=r_+$.
Thus, we can conclude that the electric Meissner effect \cite{BGR1,BGR2,BGR3,BGR4}
emerges for the maximally-charged dilaton black hole, regardless of the dilatonic
charge of the point charge.

Incidentally,  in
Appendix~\ref{AA}, we show approximate analytical expressions for $W$ and $\phi$
separately near the horizon of the maximally-charged black hole.
\section{Summary and discussion}
\label{conclusion}

In this paper we have studied the perturbative approach to the static
configuration of the charged dilaton black hole and a massive particle with
electric and dilatonic charges.
We found that the compatible condition of the coupled equation coincides with the
static condition of the massive test particle with charges in the background of a
charged dilaton black hole. 
The exact analytical expression of the linear perturbation of fields has been
found for the case with $\beta=\alpha$. We also found that the component of the
electric field normal to the outer horizon tends to vanish as the limit
$r_-\rightarrow r_+$ and the flux lines are expelled in the limit. This result
is independent of the values of $\alpha$ and $\beta$. This is the ``electric
Meissner effect'', which has been found for the Reissner--Nordstr\"om black hole
\cite{BGR1,BGR2,BGR3,BGR4}. 

We performed analytic study of the system in the present paper, but to obtain
solutions for the general case with $\beta\ne\alpha$ and $r_-\ne r_+$,
we need a numerical calculation. 
Contrary to the general case, the specific system with $\beta\approx\alpha$
and $r_-\approx r_+$ may be investigated through the
perturbative calculation on the exact multi-black hole solution.
We will try to carry out the calculations in the future work.

A straightforward extension of the theoretical investigation can be thought on the
system including magnetically charged dilaton black hole \cite{GM,GHS,KE,KM},
and on the generalized Einstein--Maxwell-scalar system described by
\cite{ABHRS,RP2,CHM,YQG,BBL}, and on the system of nonlinearly charged black holes
\cite{HPPM,RCBKPH,HPP,PHPH,Dehghani1,Dehghani2,Dehghani3,DPR,PR1,PR2,RP1,PR3,PR4,DS},
etc. We are going to address these subjects in our further work.

\appendix

\section{Approximate analytical expressions for $W$ and $\phi$ near the
horizon of the maximally-charged dilaton black hole ($r_-=r_+$)}\label{AA}

We consider an approximation valid for $\beta-\alpha\ll 1$.
First, we assume that the contributions of the terms including
$\alpha W'(r)+2\phi'(r)$ in (\ref{74}) and (\ref{75}) are small. 
Then, we assume
$\mathcal{W}(r,\theta)=\sm
W(r)Y(\theta)=\sm
[W_0(r)+W_1(r)]Y(\theta)=\mathcal{W}_0(r,\theta)+\mathcal{W}_1(r,\theta)$
and
$\tilde{\phi}(r,\theta)=\sm \phi(r)Y(\theta)=\sm
[\phi_0(r)+\phi_1(r)]Y(\theta)=\tilde{\phi}_0(r,\theta)+
\tilde{\phi}_1(r,\theta)$, where $W_1$ and $\phi_1$ are small.
Thus, we can obtain
\begin{eqnarray}
\mathcal{W}_0(r,\theta)&=&\frac{2m
\left(1-\frac{r_+}{b}\right)^{\frac{\alpha(\beta-\alpha)}{1+\alpha^2}}\left(r-r_+\right)}{r\sqrt{\left(r-r_+\right)^2+
\left(b-r_+\right)^2-2
\left(r-r_+\right)\left(b-r_+\right)\cos\theta}}
\,,
\\
\tilde{\phi}_0(r,\theta)
&=&-\frac{\beta
m\left(1-\frac{r_+}{b}\right)^{\frac{\alpha(\beta-\alpha)}{1+\alpha^2}}
\left(r-r_+\right)}{r\sqrt{\left(r-r_+\right)^2+
\left(b-r_+\right)^2-2
\left(r-r_+\right)\left(b-r_+\right)\cos\theta}}\,.
\end{eqnarray}
Consequently, we find
\begin{equation}
\alpha\mathcal{W}_0+2\tilde{\phi}_0=\frac{2m(\alpha-\beta)
\left(1-\frac{r_+}{b}\right)^{\frac{\alpha(\beta-\alpha)}{1+\alpha^2}}
\left(r-r_+\right)}{r\sqrt{\left(r-r_+\right)^2+
\left(b-r_+\right)^2-2
\left(r-r_+\right)\left(b-r_+\right)\cos\theta}}\,.
\label{A3}
\end{equation}
Admittedly, this turns out to be small in the sense of $\beta-\alpha\ll 1$.
Since the following simple expansion is well known:
\begin{equation}
\frac{1}{\sqrt{1-2tx+t^2}}=\sum_{l=0}^\infty P_l(x) t^l\,,
\end{equation}
the combination (\ref{A3}) can be expressed as
\begin{eqnarray}
\alpha\mathcal{W}_0+2\tilde{\phi}_0&=&\frac{2m(\alpha-\beta)
\left(1-\frac{r_+}{b}\right)^{\frac{\alpha(\beta-\alpha)}{1+\alpha^2}}
}{r}\sum_{l=0}^\infty \frac{(r-r_+)^{l+1}}{(b-r_+)^{l+1}}P_l(\cos\theta)
\quad(r<b)\,,\label{nh}
\\
\alpha\mathcal{W}_0+2\tilde{\phi}_0&=&\frac{2m(\alpha-\beta)
\left(1-\frac{r_+}{b}\right)^{\frac{\alpha(\beta-\alpha)}{1+\alpha^2}}
}{r}\sum_{l=0}^\infty \frac{(b-r_+)^{l}}{(r-r_+)^{l}}P_l(\cos\theta)
\quad(r>b)\,.
\end{eqnarray}
These can be used to find the next order, but the differential equations
(\ref{74}) and (\ref{75}) are difficult to solve exactly. Therefore, we
demonstrate an approximate analytical approach at
$r-r_+\ll r_+$. Using (\ref{nh}), the equations (\ref{74}) and
(\ref{75}) can be read in this order as
\begin{eqnarray}
& &(rW_1)''
-\frac{l(l+1)}{(r-r_+)^2}rW_1
\nonumber \\
&=&-\frac{2\alpha}{1+\alpha^2}\frac{2m(\alpha-\beta)r_+
\left(1-\frac{r_+}{b}\right)^{\frac{\alpha(\beta-\alpha)}{1+\alpha^2}}
}{r-r_+}
\sqrt{\frac{4\pi}{2l+1}}\partial_r\frac{(r-r_+)^{l+1}}{r(b-r_+)^{l+1}}\,,
\\
& &(r\phi_1)''
-\frac{l(l+1)}{(r-r_+)^2}r{\phi}_1\nonumber \\
&=&
-\frac{1}{1+\alpha^2}
\frac{2m(\alpha-\beta)r_+
\left(1-\frac{r_+}{b}\right)^{\frac{\alpha(\beta-\alpha)}{1+\alpha^2}}
}{r-r_+}\sqrt{\frac{4\pi}{2l+1}}\partial_r\frac{(r-r_+)^{l+1}}{r(b-r_+)^{l+1}}\,.
\end{eqnarray}
Since $r-r_+\ll r_+$, we adopt the approximation
\begin{equation}
\partial_r\frac{(r-r_+)^{l+1}}{r(b-r_+)^{l+1}}\approx
\frac{(l+1)(r-r_+)^{l}}{r(b-r_+)^{l+1}}
\end{equation}
and we get
\begin{eqnarray}
& &(rW_1)''
-\frac{l(l+1)}{(r-r_+)^2}rW_1
\nonumber \\
&=&-\frac{4\alpha m(\alpha-\beta)}{1+\alpha^2}{
\left(1-\frac{r_+}{b}\right)^{\frac{\alpha(\beta-\alpha)}{1+\alpha^2}}
}(l+1)
\sqrt{\frac{4\pi}{2l+1}}\frac{(r-r_+)^{l-1}}{(b-r_+)^{l+1}}\,,
\\
& &(r\phi_1)''
-\frac{l(l+1)}{(r-r_+)^2}r{\phi}_1\nonumber \\
&=&
-\frac{2m(\alpha-\beta)}{1+\alpha^2}
{
\left(1-\frac{r_+}{b}\right)^{\frac{\alpha(\beta-\alpha)}{1+\alpha^2}}
}(l+1)\sqrt{\frac{4\pi}{2l+1}}\frac{(r-r_+)^{l-1}}{(b-r_+)^{l+1}}\,.
\end{eqnarray}
Finally, the solutions for these approximate equations:
\begin{eqnarray}
W_1&=&-\frac{4\alpha m(\alpha-\beta)}{1+\alpha^2}{
\left(1-\frac{r_+}{b}\right)^{\frac{\alpha(\beta-\alpha)}{1+\alpha^2}}
}\frac{l+1}{2l+1}
\sqrt{\frac{4\pi}{2l+1}}\frac{(r-r_+)^{l+1}}{(b-r_+)^{l+1}}
\ln\frac{r-r_+}{C(b-r_+)}\,,
\\
\phi_1&=&
-\frac{2m(\alpha-\beta)}{1+\alpha^2}
{\left(1-\frac{r_+}{b}\right)^{\frac{\alpha(\beta-\alpha)}{1+\alpha^2}}
}\frac{l+1}{2l+1}\sqrt{\frac{4\pi}{2l+1}}\frac{(r-r_+)^{l+1}}{(b-r_+)^{l+1}}
\ln\frac{r-r_+}{C(b-r_+)}\,,
\end{eqnarray}
where $C$ is the undetermined constant in the present approach.
However, we can determine that the solutions in this order is regular at $r=r_+$.%
\footnote{The summed expressions $\mathcal{W}_1$ and $\tilde{\phi}_1$
are possible but they are not needed for this time, so we omit them.}
Notice that the electric field around the maximally-charged black hole with a
point charge does not change even for $\beta\ne\alpha$.


\bibliographystyle{apsrev4-1}


\end{document}